%% file: main.tex
\documentclass[conference]{IEEEtran}
\IEEEoverridecommandlockouts

\input{macros}

\begin{document}

\title{Let's Talk About It: Making Scientific Computational Reproducibility Easy


}

\author{
\IEEEauthorblockN{Lázaro Costa}
\IEEEauthorblockA{\textit{Faculty of Engineering, University of Porto}\\
\& \textit{INESC TEC, Portugal} \\
lazaro@fe.up.pt}
\and
\IEEEauthorblockN{Susana Barbosa}
\IEEEauthorblockA{\textit{INESC TEC, Portugal}\\
susana.a.barbosa@inesctec.pt}
\and
\IEEEauthorblockN{Jácome Cunha}
\IEEEauthorblockA{\textit{Faculty of Engineering, University of Porto}\\
\& \textit{HASLab/INESC TEC, Portugal} \\
jacome@fe.up.pt}
}

\maketitle

\begin{abstract}

Computational reproducibility of scientific results, that is, the execution of a computational experiment (e.g., a script) using its original settings (data, code, etc.), should always be possible.
However, reproducibility has become a significant challenge, as researchers often face difficulties in accurately replicating experiments due to inconsistencies in documentation, setup configurations, and missing data. This lack of reproducibility may undermine the credibility of scientific results.

To address this issue, we propose a conversational, text-based tool that allows researchers to easily reproduce computational experiments (theirs or from others) and package them in a single file that can be re-executed with just a double click on any computer, requiring the installation of a single widely-used software. Researchers interact with the platform in natural language, which our tool processes to automatically create a computational environment able to execute the provided experiment/code. 
Our proposal simplifies the reproducibility process, making it easier for researchers to verify and build upon previous work. This approach directly addresses the reproducibility crisis by providing a user-friendly system that ensures consistency in computational research.

We conducted two studies to evaluate our proposal. In the first study, we gathered qualitative data by executing 18 experiments from the literature. Although in some cases it was not possible to execute the experiment, in most instances, it was necessary to have little or even no interaction for the tool to reproduce the results.

We also conducted a user study comparing our tool with an enterprise-level one. During this study, we measured the usability of both tools using the System Usability Scale (SUS) and participants' workload using the NASA Task Load Index (TLX). The results show a statistically significant difference between both tools in favor of our proposal, demonstrating that the usability and workload of our tool are superior to the current state of the art.
\end{abstract}



\section{Introduction}\label{sec:interfaceIntro}
Reproducibility has long been recognized as a cornerstone of scientific progress, although it may have different means in different scientific fields \cite{reproducibility2019report}. In our work, we focus on computation reproducibility or simply reproducibility, that is, ``\textit{obtaining consistent results using the same input data; computational steps, methods, and code; and conditions of analysis}'' \cite{reproducibility2019report}. 

However, in computational research, reliably re-executing code to achieve consistent results remains a persistent challenge \cite{Ivie2018}. Researchers in fields like computer science, machine learning, bioinformatics, or computational physics often struggle to replicate experiments due to discrepancies in software environments, incomplete documentation, and a lack of transparency in data usage \cite{costa2024Rep,Ivie2018,reproducibility2019report}. Even minor variations in hardware or software configuration can lead to dramatically different results, making the verification of published findings at least challenging \cite{reproducibility2019report}. This reproducibility crisis has raised concerns about the reliability and credibility of scientific outcomes as reported in the media already more than 10 years ago \cite{crisis}.

The lack of reproducibility is particularly problematic in computational research because experiments are often complex, involving intricate workflows that combine multiple software tools, different programming languages (PLs), data sources, and environment variables \cite{Stodden2016}. Furthermore, traditional methods of documenting experiments, such as written descriptions or manually recorded steps, are prone to human error and omission. As a result, even with detailed publications, critical elements of an experiment may be lost or miscommunicated (we show an example in \cref{sec:example}), making reproducibility difficult or impossible \cite{Hunter2017}.

To aid researchers, multiple tools have been proposed. However, some of these tools have limited support for different PLs (e.g., Binder~\cite{Matthias2018binder}, Code Ocean~\cite{codeocean}, FLINC~\cite{Ahmad2022Flinc}, RenkuLab~\cite{ramakrishnan2023renku} and WholeTale~\cite{Brinckman2019}), can only be executed in particular operating systems (e.g., FLINC~\cite{Ahmad2022Flinc}, Provenance-To-Use (PTU)~\cite{Pham2013PTU}, Sciunit~\cite{Ton2017SciUnits}, and Reprozip~\cite{Chirigati2020}), are outdated or unavailable (e.g., CARE~\cite{Janin2014}, Code, Data, Environment (CDE)~\cite{Philip2011}, Encapsulator~\cite{Pasquier2018}, PARROT~\cite{thain2005parrot}, Prune~\cite{Ivie2016Prune}, reprozip-jupyter~\cite{reprojupyter}, ResearchCompendia~\cite{Stodden2015ResearchCompendia} and SOLE~\cite{malik2014sole} and Umbrella~\cite{Meng2015Umbrella}), or have user interfaces that require knowledge many researchers don't have, especially if we consider researchers not related to the computer science field (e.g., Binder~\cite{Matthias2018binder}, Code Ocean~\cite{codeocean}, RenkuLab~\cite{ramakrishnan2023renku} and WholeTale~\cite{Brinckman2019}) \cite{costa2024Rep}. See \cref{sec:related} for more details about related work.

To address these limitations, we introduce \tool, a conversational tool designed to simplify the process of computational reproducibility by allowing researchers to interact with a system using natural language. As detailed in \cref{sec:sciconv}, \tool automates the identification of software dependencies, infers execution environments, and assists users in troubleshooting errors, reducing the manual effort required to configure reproducibility packages. 

This work examines whether a conversational interface can enhance the usability and accessibility of reproducibility tools for researchers from different fields, which we explore through the following Research Question (RQ):\\

\noindent
\textbf{RQ: Can a conversational interface improve the usability and accessibility of computational reproducibility tools for researchers across different fields?}\\

To answer this RQ, we designed and evaluated \tool, aligning with recent developments in LLMs to facilitate researcher-tool interactions. Our evaluation consists of two complementary studies. The first, presented in \cref{sec:interfaceQualitative}, assesses \tool's ability to automate reproducibility using a curated dataset of 18 experiments previously defined~\cite{Costa2025REP}. The results indicate that most experiments were successfully reproduced with minimal user input, demonstrating \tool’s ability to infer execution environments and dependencies automatically. However, limitations were identified, particularly in handling experiments involving external databases.

To evaluate the usability and the workload required by our proposal, we executed a comparative study with a professional online platform, \co, through a user study involving 21 researchers from diverse fields. We measured usability and workload using the SUS \cite{sus} and NASA TLX \cite{tlx} questionnaires, finding statistically significant differences favoring \tool. These results confirm that \tool not only improves usability but also reduces the cognitive workload associated with computational reproducibility compared to \co. We provide more details in \cref{sec:RQ1.5-interface}.

We present conclusions and future work in \cref{sec:conclusions_interface}.

In the next section, we introduce a running example to make the challenges of reproducing more evident. The remainder of this work introduces \tool and presents its design, evaluation, and findings. We conclude with a discussion on its broader implications and potential improvements (\cref{sec:conclusions_interface}).

\section{An Example of Reproducibility}\label{sec:example}

In this section, we present an example of a computational experiment used in a scientific publication \cite{Chen2022}. The source code can be accessed at \url{https://github.com/OpsPAI/ADSketch}. This is actually a good example where the authors provide information about the code, how to use it, and its dependencies (although this is provided in a \texttt{requirements.txt} file, which may not be immediately clear to everyone).
We start by presenting our experience with trying to reproduce this code using a professional tool (\cref{sec:usingCodeocean}), followed by the use of our tool (\cref{sec:usingSciconv}).

\subsection{Using a Professional Tool}\label{sec:usingCodeocean}

To reproduce this experiment, we used \textit{Code Ocean}\footnote{Available at \url{https://codeocean.com}.}, which is a widely-used enterprise-level online platform for computational reproducibility.
After creating a new project and uploading all the code, we defined the necessary parameters to execute the code, as described by the experiment's authors. Thus, we defined the language as Python 3.6.3, selected the library dependencies, and defined the execution file. Finally, we tried to execute our experiment, but it failed, providing us an error because Python's package manager (\texttt{pip} in this case) was not able to install the dependencies. \cref{fig:codeoceanerror} at \cref{app:image} illustrate the error shows.

To solve the problem, we manually edited the environment that Code Ocean automatically created based on our previous choices and added a command to update \texttt{pip} as suggested. This is actually not incentivized by the platform as this required editing a \textit{dockerfile}\footnote{A dockerfile is the file used by Docker to create computational environments and is a highly technical file used by software engineers -- see \url{https://docs.docker.com/reference/dockerfile/} for more details.}, a file probably not every research is comfortable editing. After this step, the experiment's code started to execute. However, we got a new error: ``\texttt{ModuleNotFoundError: No module named `tqdm'}''. This seems to indicate a necessary library is missing from the dependencies list. This can happen for multiple reasons: maybe the authors forgot to add it, maybe the library was part of some other library of Python version, etc. This also shows that even the code provided with great care may still miss vital information for its execution. After adding this dependency, we were finally able to execute the experiment.

We try to illustrate with this example that even a ``simple'' Python script, for which the authors provide documentation, and even using enterprise-level software, may be challenging to execute exactly as the authors defined. Note that, being computer science researchers, we are comfortable with the steps we took to solve the recurring problems, but this is not the case for many researchers less familiar with such low-level details of generic execution environments.

\subsection{Using \tool}\label{sec:usingSciconv}
These are the steps of our approach, detailing what happens during execution in a simple way. We have used the same example code with our tool.

In this case, the researcher needs to provide a \textit{zip} file with all the content (code, datasets, etc.). The tool immediately starts to automatically identify executable files and detect configuration files needed to set up the project.

During this process, the tool provides messages to the researcher explaining which phase it is in so the user does not feel lost. Following this, the tool asks the researcher to insert the command needed to execute the project and infers the PLs involved, library dependencies, etc.

Next, the tool builds a Dockerfile without requiring user input. It uses the information inferred in the previous step to automatically generate a Dockerfile containing all the necessary details to execute the project.

Then, the tool builds the Docker image, also without user input. It automatically builds the Docker image based on the Dockerfile generated earlier, incorporating all necessary dependencies and configurations. If an error occurs during this build process, the workflow transitions to the "Wait For Chat Interaction" step for troubleshooting and further input.

Afterward, the tool runs the container without needing any user input. It uses the Docker image created earlier, along with the command(s) provided by the researcher, to execute the project. If an error occurs during execution, the process will automatically move to the "Wait For Chat Interaction" step for further guidance.

The tool then packages all the necessary information into a zip file, including the environment setup—such as dependencies and PLs used. Additionally, the zip file contains both a batch script (for Windows) and a bash script (for Unix-based systems) that can be executed to reproduce the same results achieved in the previous step. This ensures cross-platform compatibility, as the file can be run on any operating system with only one requirement: Docker must be installed.

Finally, the tool notifies the researcher that the procedure has been successfully completed. If any issues arise during the process, the tool enters the ``Wait For Chat Interaction'' stage, where it interacts with the researcher to identify the problem's source and collaborates to find a solution. If the researcher responds positively and indicates that the problem can be resolved, the tool can return to the step immediately before where the issue occurred, with options to revisit some previous stages.


\section{Related Work}\label{sec:related}


We separate our work into two different areas. On the one hand, we discuss other approaches for reproducibility in \cref{sec:repro}; on the other hand, we discuss approaches related to artificial intelligence assistants in \cref{sec:aiassist}.

\subsection{Reproducibility}\label{sec:repro}

In a previous study~\cite{costa2024Rep}, the authors conducted a comprehensive literature review on the available tools aimed at assisting researchers from various disciplines in addressing the challenges related to reproducibility. Through an initial search followed by snowballing, they identified a total of 18 tools:
Binder~\cite{Matthias2018binder}, CARE~\cite{Janin2014}, Code, Data, Environment (CDE)~\cite{Philip2011}, Code Ocean~\cite{codeocean}, Encapsulator~\cite{Pasquier2018},  FLINC~\cite{Ahmad2022Flinc}, PARROT~\cite{thain2005parrot}, 
PTU~\cite{Pham2013PTU}, Prune~\cite{Ivie2016Prune}, RenkuLab~\cite{ramakrishnan2023renku}, Reprozip~\cite{Chirigati2020}, reprozip-jupyter~\cite{reprojupyter}, ResearchCompendia~\cite{Stodden2015ResearchCompendia}, SciInc~\cite{Youngdahl2019SciInc}, Sciunit~\cite{Ton2017SciUnits}, SOLE~\cite{malik2014sole},  Umbrella~\cite{Meng2015Umbrella}, and WholeTale~\cite{Brinckman2019}.
However, after they analyzed all the tools they could find online, only 8 tools were available, and they could use, namely: Binder, Code Ocean, FLINC, PTU, RenkuLab, Reprozip, Sciunit, and WholeTale.
We now discuss these tools in five different criteria: PLs, automatic inference, automatic installation, packaging, and user interface  (UI).

\subsection*{C1: Supported PLs}
Researchers use a variety of PLs for experiments, so it's crucial that any approach can accommodate this diversity. Tools like Binder, Code Ocean, FLINC, RenkuLab, and Whole Tale support only a limited set of PLs and cannot be extended to include new ones. In contrast, ReproZip, Sciunit, PTU, and \toolbackend support an unlimited range of PLs.

\subsection*{C2: Automatically Detects Dependencies}
One of the main challenges in computational reproducibility is ensuring that a framework can identify all necessary dependencies and libraries to run code as originally executed.

Tools like FLINC, ReproZip, Sciunit, and PTU automatically track dependencies using observed provenance capture, revealing script details. In contrast, Binder, RenkuLab, Whole Tale, and Code Ocean lack full dependency detection.

\toolbackend also automatically infers dependencies, such as PLs and libraries.

\subsection*{C3: Automatically Installs Dependencies}
A key challenge in computational reproducibility is correctly installing the required libraries and dependencies.

Tools like FLINC, ReproZip, Sciunit, PTU, Binder, RenkuLab, and Whole Tale can automate this process. However, Binder, RenkuLab, and Whole Tale rely on dependencies being listed in a specification file, which can be difficult for less experienced researchers. Code Ocean requires the manual addition of all dependencies.

\toolbackend also automates the installation of all dependencies.

\subsection*{C4: Research Artifact}
The integrity and reproducibility of research depend on the availability and accessibility of key artifacts like code, datasets, and dependencies. These artifacts must be well-documented, easily accessible, and replicable across different environments.

Although tools such as Binder and RenkuLab do not support the creation of artifacts, others such as Code Ocean, FLINC, PTU, ReproZip, Sciunit and Whole Tale allow these components to be built and managed, enabling effective sharing and reproduction.

\tool simplifies artifact creation by using Docker, requiring only this one dependency. Although Docker is widely adopted, its complexity can be a barrier for typical researchers. \tool abstracts this complexity, offering a user-friendly interface aligned with researchers' needs.

\subsection*{C5: User Interface}
The usability and accessibility of a tool are crucial factors in ensuring researchers can efficiently create and reproduce experiments. A well-designed user interface simplifies complex tasks and enhances user interaction, making it easier for researchers to adopt and utilize the tool effectively.

Tools like Binder, Code Ocean, FLINC, RenkuLab and Whole Tale offer user interfaces, allowing researchers to interact with the tool through a more intuitive and visual experience. These interfaces simplify workflows and make tools accessible, although they don't always present the right workflow for researchers with limited technical knowledge.

In contrast, tools such as PTU, ReproZip, and Sciunit lack a dedicated user interface, which can present challenges for researchers who are less comfortable with the command-line interface.

\tool features a conversational user interface that allows researchers to reproduce experiments through natural language interactions. This innovative approach enhances usability and aligns with researchers’ workflows, making creating and reproducing experiments more intuitive and efficient.

\subsection{Artificial Intelligence Assistants}\label{sec:aiassist}

Recent advancements in large language models (LLMs) have sparked significant interest in utilizing these models to enhance the efficiency of various approaches, including the generation of user interface (UI) elements. NL2INTERFACE~\cite{NL2INTERFACE} has been developed to leverage natural language queries for creating interactive multi-visualization interfaces. This tool enables users to input queries, automatically generating interfaces without the need for programming knowledge or specialized tools. Similarly, other tools like LIDA~\cite{LIDA} allow users to create data visualizations directly from natural language inputs. DynaVis~\cite{Priyan2024} employs an LLM to synthesize interactive visualizations with dynamically generated UI widgets tailored to the visualizations, enabling direct manipulation of their properties. Another tool, PDFChatAnnotator~\cite{Tang2024}, demonstrates the significant role that LLMs play in enhancing data annotation processes. This innovative tool addresses the challenge of extensive unannotated data in PDF catalogs by facilitating a collaborative environment between humans and LLMs.

Researchers have highlighted the critical relevance of LLMs in enhancing user experiences in software assistance contexts~\cite{Khurana2024}. It reveals that while LLMs like chatGPT have been widely adopted for various workflows, users frequently encounter dissatisfaction due to the models' inability to accurately interpret user intentions and provide contextually relevant responses. In the same way, Chiang et al.~\cite{Chiang2024} investigate various interaction styles of the ``devil's advocate'' and assess their influence on group dynamics and perceptions, revealing that LLMs can significantly enhance the quality of discussions and outcomes in collaborative settings.

In alignment with these developments, our approach also utilizes LLMs to improve interactions with researchers, facilitating the reproduction of experiments in a more intuitive and efficient manner.


\section{\tool}\label{sec:sciconv}
In this section, we provide a comprehensive overview of \tool, covering its core design principles (\cref{sec:interface-goals}), architectural framework and components (\cref{sec:arch}), implementation details (\cref{sec:implementation}), and the user interaction model that enables the easy reproducibility of workflows (\cref{sec:interaction}).

\subsection{Design Goals}\label{sec:interface-goals}
Reproducibility is challenging, even for computer science researchers, but even more so for others
less used to designing and developing software. Thus, we wanted to design a tool that was simple to use, 
guiding the user through the process, and that could cope with possible problems during the process.

\begin{description}
        \item[\textbf{Simple}]\hspace*{0.5em} The tool must be simple to use. This is a very abstract concept, but after doing some experiments with ``traditional’’ Graphical User Interface (GUI) (web-based, online with tools such as \co\footnote{Available online at: \url{https://codeocean.com/}} or Binder~\cite{Matthias2018binder}, we recognized they still require quite some interaction with researchers. Given the widespread adoption of artificial intelligence (AI) chat systems across various domains, we chose to explore a conversational approach. Our tool should be as intuitive as having a natural conversation about reproducibility needs.

        \item[\textbf{Assistive}] \hspace*{1em}
        Traditional GUI requires users to navigate menus, forms, and settings, which can be unintuitive. AI-driven assistants, on the other hand, provide a more natural and adaptive interaction, making them more effective for guiding users through complex workflows. Thus, our tool should assist and guide the user through the process.

	\item[\textbf{Error supportive}]\hspace*{4.5em} Burnett in her keynote at the conference VL/HCC 2023 \cite{10305657}, emphasized that it is not sufficient to create systems that work from one end to the other, but they must also cope with problems along the user's journey. Indeed, users often face problems when using a tool, and in many cases, there is no help. An example is the situation we described when using \co in \cref{sec:example}, when we got an error and were not provided with any kind of help on how to move forward. Thus, our tool should assume researchers will encounter problems and should be prepared to help them recover.
\end{description}

\subsection{Design and Architecture}\label{sec:arch}
\subsubsection{Interaction Model Between the User, Our Tool's Core, and an Assistant}

We designed our tool with extensibility in mind, enabling it to adapt and support a variety of related processes beyond its initial scope. While the tool's current focus is on guiding users through the creation of a reproducibility package, its architecture is versatile enough to accommodate different workflows or entirely new processes with varying steps. This adaptability ensures that the tool remains relevant and valuable across diverse research contexts.

\begin{figure}
    \centering
    \includegraphics[width=1\linewidth]{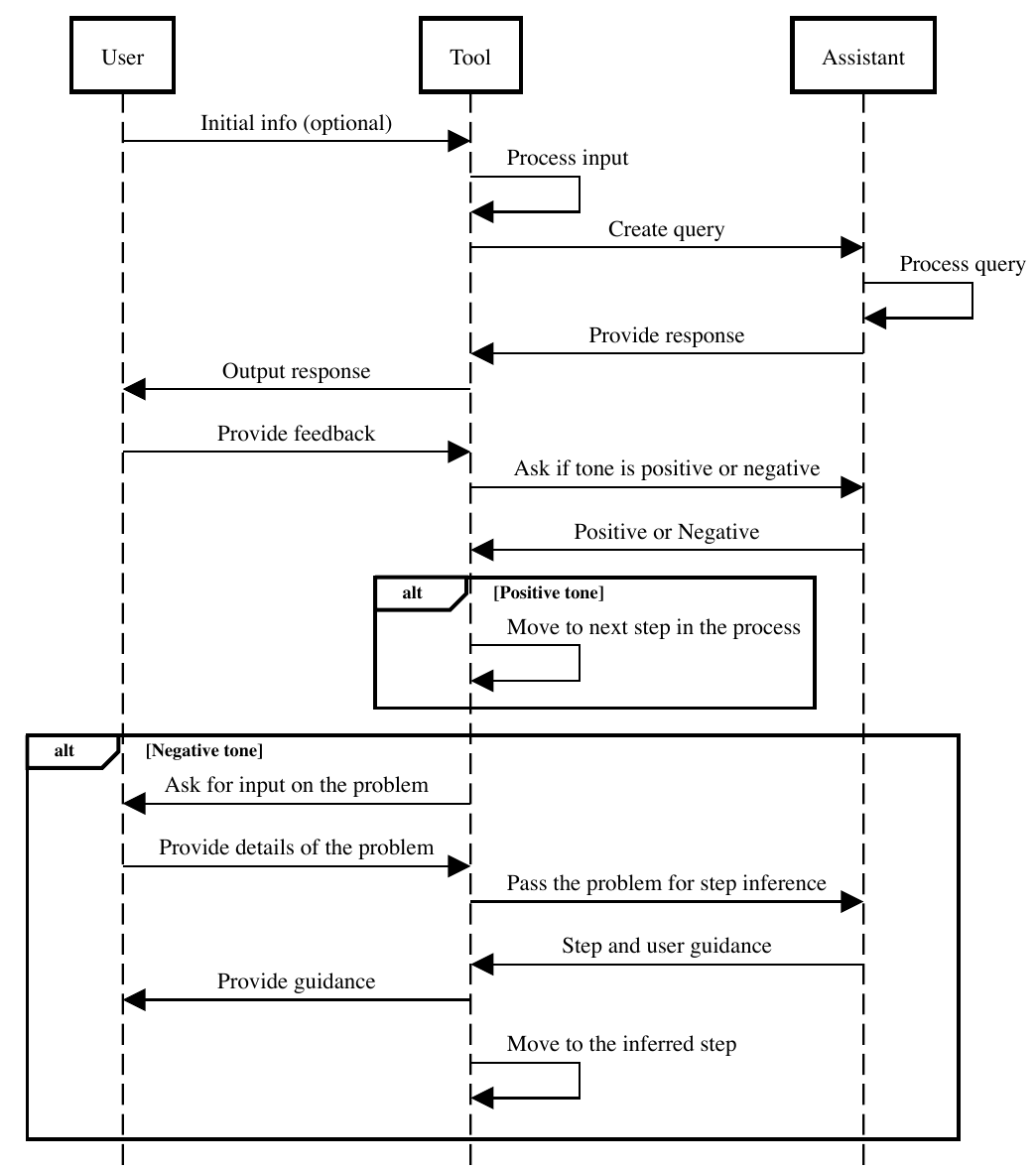}
    \caption{A sequence diagram of user, tool, and conversation system}
    \label{fig:interaction}
\end{figure}

To illustrate the interaction dynamics, \cref{fig:interaction} presents a sequence diagram featuring three key lifelines: the user/researcher, the core of our tool, and the supporting conversational system, which we term the assistant. The assistant, which can be powered by an AI chat model or an alternative system, plays a pivotal role in facilitating smooth communication and enhancing the user experience.

At each step, the user has the option to provide information to the tool, although this is not always required. Many steps can be performed autonomously by the system without user input. Ideally, the entire process would be fully automated, eliminating the need for user interaction altogether.

When user input is provided, the tool processes it and formulates a query for the assistant. The assistant, in turn, processes the query and returns a response to the tool. The tool then uses this response to generate an appropriate reply for the user. 
In some cases, the tool may directly relay the assistant’s response, while in others, it may interpret and refine the response to provide more tailored and context-specific guidance.

For example, if a researcher specifies that Python 3.10 is required, the tool would prompt the assistant to include this version information in the configuration file being generated. The assistant would then return the full set of configuration details, which the tool would present to the user for validation. 

After receiving the tool's response, the user can provide feedback, which is particularly flexible in this conversational interface. The user can input any message, which is then processed by the tool. To interpret the feedback, the tool analyzes the message's tone by querying the assistant to determine whether it is positive or negative.

Positive feedback allows the process to continue, while negative feedback—such as missing dependencies—triggers a guided troubleshooting interaction to resolve the issue. In such cases, the tool prompts the user for more details about the issue. This additional input is then sent to the assistant, enabling it to infer which step in the process the user is referencing.

The assistant, which has been designed with awareness of the process's various stages, uses this information to determine the most relevant step and suggests how to address the issue. The tool then revisits the inferred step, offering updated guidance or adjustments based on the user's input.

This interaction model is highly adaptable and can be generalized for a wide range of processes. However, each specific implementation of the model must be tailored to align with the particular process, the type of assistant used, and the ultimate objectives of the tool.

\subsubsection{The Current Process}
The current process is structured into multiple steps, each designed to facilitate and simplify the user experience. While most of these steps are automated and do not require user input, we provide specific details for steps where user interaction is necessary.
\begin{description}


    \item[\textbf{Project Location:}] \hspace*{4.5em} This step requires user input, specifically the upload of a zip file containing the project.

    \item[\textbf{Find Project Files:}] \hspace*{5em}  Our approach automatically identifies executable files and detects configuration files needed to set up the project.

    \item[\textbf{Parameters to Use:}] \hspace*{5.5em} This step requires the participation of the researcher, who must provide the command(s) needed to execute the project. These commands will be executed in the following steps, following the order specified.

    \item[\textbf{Find Configurations:}] \hspace*{6.5em} Our approach automatically infers all the PLs and dependencies required to run the project.

    \item[\textbf{Build Dockerfile:}] \hspace*{4.5em}  \tool leverages the information inferred in the previous step to automatically generate a dockerfile containing all the necessary details to execute the project.

    \item[\textbf{Build Docker Image:}] \hspace*{6em} The tool automatically builds the Docker image\footnote{A Docker image is a standardized package containing all the necessary binaries, configuration files, dependencies, etc., that allow Docker to run the experiment. This can be seen as a very small Virtual Machine (VM). More details are available at \url{https://docs.docker.com/get-started/docker-concepts/the-basics/what-is-an-image/}.} based on the dockerfile generated in the previous step, incorporating all the necessary dependencies and configurations. This ensures that the project can be reliably executed across different environments. If an error occurs during the build process, the workflow transitions to the `Wait for Chat Interaction' step for troubleshooting and further input.

    \item[\textbf{Run Container:}] \hspace*{4em} Our approach uses the Docker image created in the previous step, along with the command(s) provided by the researcher in the `Parameters To Use' step, to execute the project. If an error occurs during execution, the process will automatically move to the `Wait For Chat Interaction' step for further guidance.

    \item[\textbf{Research Artifact:}] \hspace*{5em} \tool packages all the necessary information into a zip file, including the environment setup, such as dependencies and PLs used. Additionally, the zip file contains both a batch script (for Windows) and a bash script (for Unix-based systems) that can be executed to reproduce the same results achieved in the previous step. This functionality ensures cross-platform compatibility, as the file can be run on any Operating System (OS) with only one requirement: Docker must be installed. 

    \item[\textbf{Completed:}] \hspace*{2em} Our approach notifies the researcher that the procedure has been successfully completed.

    \item[\textbf{Wait For Chat Interaction:}] \hspace*{9em} This stage requires the participation of the user. Our approach interacts with the researcher to identify the source of the problem and collaborates with them to find a solution. If the researcher responds positively and indicates that the problem can be solved, our approach will return to the step immediately ahead of where the problem occurred. 

\end{description}

\subsubsection{Architecture}

\begin{figure}
    \centering
    \includegraphics[width=01\linewidth]{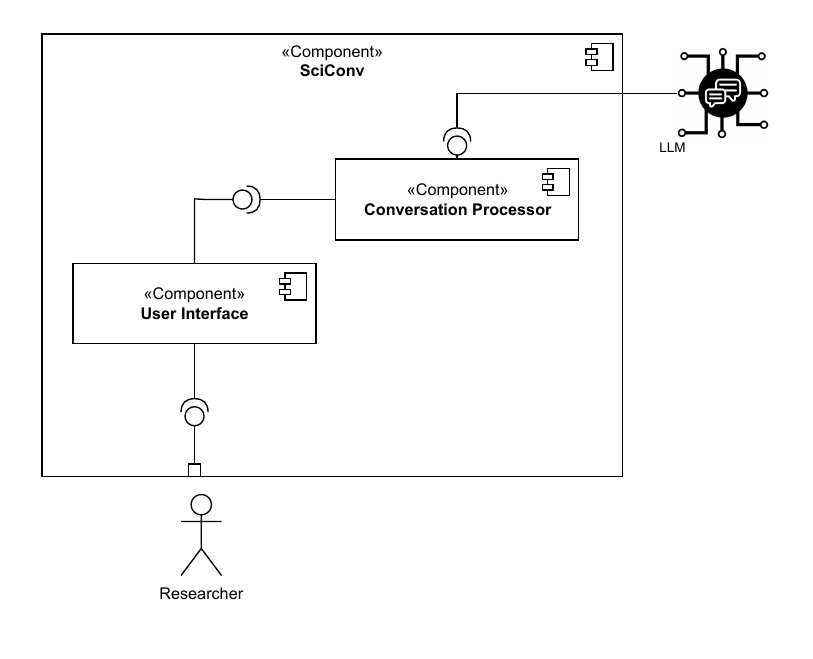}
    \caption{Architecture diagram of \tool}
    \label{fig:arch}
\end{figure}

Based on the design presented, we built the tool’s architecture, illustrated in \cref{fig:arch}.

The architecture is designed to be simple yet effective, comprising two key components:
\begin{description}
    \item[\textbf{User Interface (UI) Component}] \hspace*{11em} This component serves as the primary point of interaction for the researcher. It offers an intuitive and user-friendly interface that allows researchers to input queries, receive responses, and manage the flow of communication. The UI is designed to ensure researchers can easily engage with the system without needing deep technical knowledge. It acts as a bridge between the researcher and the underlying core component.

    \item[\textbf{Conversation Processor Component}]  \hspace*{13em}The conversation processor serves as the core of the architecture and is responsible for managing interactions with the underlying LLM. It interfaces directly with the LLM and processes the data coming from both the user and the model. This component ensures that the researcher's queries are appropriately formatted and sent to the LLM for processing and that the LLM's responses are refined, filtered, and contextualized before being passed back to the researcher. Acting as a mediator ensures that the flow of information remains coherent and relevant. 

    \item[\textbf{LLM Component}] \hspace*{5em} We currently use the GPT-4-Turbo model in our approach.
\end{description}

\subsection{Implementation}\label{sec:implementation}
We developed a web-based chat interface to simplify researcher and assistant communication. The front-end is built using Angular~\footnote{\url{https://angular.dev/}}, which enables the creation of a highly dynamic and responsive UI. This component efficiently manages user inputs, renders assistant responses, and provides an intuitive user experience throughout the interaction.

We reuse the logic part developed in previous work~\cite{costa2025backend}. 
Finally, we integrated the OpenAI Application Programming Interface (API) using the GPT-4-Turbo model as the core assistant. This model processes natural language queries and generates intelligent responses, which are passed back through the system for display to the researcher.

The complete implementation is available as an open-source project on GitHub\footnote{\tool project available at~\url{https://github.com/lazarocosta/SciConv}}.

\subsection{User Interaction}\label{sec:interaction}
The interaction with \tool takes place through a web-based chat interface designed to guide researchers through the reproducibility process with minimal technical overhead. Researchers communicate with the tool via a text box, where they can input commands, provide necessary details, or seek assistance. In response, the tool displays messages in dedicated response boxes, offering real-time feedback and instructions. Some response boxes also feature a ``See Examples'' button, allowing researchers to access additional guidance and sample inputs, helping them interact more effectively with the tool.

Once the researcher uploads a compressed zip file containing the project’s resources—such as code, datasets, and additional configuration files—\tool automatically identifies executable files and detects relevant configuration settings. Throughout the process, the tool maintains communication with the researcher, providing status updates at each step to ensure clarity and prevent uncertainty.

To execute the experiment, the tool prompts the researcher to specify the necessary command. Using this information, \tool infers the PLs, library dependencies, and execution environment details required for successful reproduction. After gathering these details, the tool automatically generates a Dockerfile, ensuring that all configurations are correctly defined without requiring further user input.

Following the Dockerfile creation, the tool proceeds to build a Docker image that encapsulates all dependencies and configurations. If any issues arise during this process, the system transitions to the ``Wait for Chat Interaction'' step, prompting the researcher to provide additional information or resolve missing dependencies. This interactive mechanism allows for quick troubleshooting and ensures that the execution environment remains consistent.

Once the Docker image is successfully built, the tool runs the experiment by executing the project within a container, using the previously specified commands. The results are displayed to the researcher, who can verify the output. If execution errors occur, the tool re-engages in an interactive session, guiding the researcher through potential fixes and allowing them to adjust configurations as needed.

To facilitate reproducibility and sharing, \tool packages the complete execution environment into a structured zip file. This package includes the computational setup, such as dependencies and language configurations, along with both a batch script (for Windows) and a bash script (for Unix-based systems). These scripts allow the experience to be easily re-executed of the experiment on any machine that has Docker installed, ensuring cross-platform compatibility and eliminating the need for additional manual setup.

At the end of the process, the tool notifies the researcher that the experiment has been successfully packaged. If any unresolved issues remain, the system re-engages with the researcher to identify the problem’s source and collaborate on a solution. If necessary, the tool allows users to revisit previous steps, making adjustments to the environment or execution parameters before finalizing the reproducibility package.

\section{Qualitative Evaluation}\label{sec:interfaceQualitative}
In this section, we present a qualitative analysis of our tool's performance using a curated dataset of experiments previously defined~\cite{Costa2025REP}. 

This initial study aims to gather qualitative insights into the functionality and adaptability of \tool when applied to experiments from diverse research domains, encompassing varying requirements such as different PLs, datasets, and dependencies.

\subsection{Study Design}
We utilized our tool to execute all 18 experiments from the curated dataset, which encompasses a broad spectrum of real-world scenarios as defined in previous work~\cite{Costa2025REP}. This dataset includes experiments spanning various domains, such as medicine and software engineering, often incorporating multiple PLs within the same experiment. By testing against such a diverse range of experiments, we aim to demonstrate the robustness and versatility of our tool in handling varied requirements.

It is important to note that this study focuses solely on assessing the tool’s technical capabilities and adaptability to different experiment types. We do not evaluate user interaction in this phase, as that aspect is reserved for the subsequent user-centered evaluation described in \cref{sec:controlled}.

\subsection{Results}
From the 18 experiments, twelve (E1, E2, E3, E5, E6, E11, E12, E13, E15, and E16) were executed requiring just one interaction, namely for us to define how the experiment should be executed (\eg, \texttt{python script.py}). This means that after we uploaded the zip file with all the experiment's content, our tool was able to automatically infer all the necessary configurations and execute the provided code. In each case, the code ran error-free, successfully producing results. Subsequently, we verified that the outputs matched the expected outcomes, confirming their correctness.
    
For five experiments (E8, E9, E10, E17, and E18), we were required to provide more information.
In one case (E8), the execution of the code produced an error (\texttt{No module named `chardet'}). Based on this, our tool asked the researcher ``\textit{What might have caused this unexpected result?}'' We wrote ``\textit{I want to add chardet dependency}'' as the error seems to provide clues about a missing dependency. After this interaction, the experiment was executed promptly.
For E9, E10, E17, and E18, the situation was similar but with different dependencies.

Another experiment requires the use of databases (E4), which we currently do not support. Although E13 also uses a database, it is a database that only requires a Python library (SQLite) and not necessarily an external database management system, like PostgreSQL, which is required by E4.


One of the experiments (E7) has too many files for our current implementation to handle.
This experiment has about 50 executable files (about 700 files in total), and since we extract a part of each file to send to GPT, it exceeds the amount of information we can send it.

Our \tool does not currently support Jupyter Notebook, which prevented the execution of one of the experiments (E14).

In summary, our approach successfully executed 15 of 18 experiments for which the code and data were available.

\subsection{Threats to Validity}
The validation of this process introduces several threats~\cite{Sarah2010,Cook1979,Wohlin2012} to validity that must be considered:

\paragraph{Construct Validity} The dataset used for evaluation was deliberately curated to encompass a diverse range of experiments across various domains. However, this selection may still introduce bias, as it includes only a single Jupyter Notebook experiment, one database-driven experiment (different from SQLite), and one large-scale experiment containing hundreds of files—all of which could not be successfully reproduced. This limited representation of complex or specialized scenarios might result in an overestimation of the tool's capabilities in environments where such configurations are more common. 
    
\paragraph{Internal Validity} Focusing solely on 15 out of 18 predefined experiments may not comprehensively capture the performance and applicability of our approach across a wide range of scenarios. The limited scope restricts our ability to generalize the results or assess the method's robustness in situations beyond those tested.

Additionally, the conversational nature of the tool relies heavily on the quality and completeness of user input. Miscommunication, incomplete descriptions, or incorrect inputs could affect reproducibility outcomes. However, these risks were mitigated through structured error-handling mechanisms, which allowed the tool to interactively request additional input and recover from common errors during the reproducibility process.

\paragraph{External Validity} The external validity of this study is influenced by the dataset's scope and the absence of comparative evaluations against other reproducibility tools. While the dataset represents a variety of domains, certain scenarios—such as experiments with highly complex dependencies, distributed systems, or domain-specific tools—remain underrepresented. Consequently, the findings may not generalize to all research fields or computational setups, particularly in cases involving these underrepresented configurations.

\paragraph{Reliability} The structured nature of the dataset, along with consistent procedures for testing and recording outcomes, provides a solid foundation for reproducibility. However, as the tool leverages conversational interactions, outcomes may vary based on the phrasing or clarity of user input. While the tool's error-handling and iterative feedback mechanisms mitigate this variability, future work could introduce more standardized input templates to ensure greater consistency in results.

Additionally, all experiments were executed under controlled conditions using a specific version of the tool and its underlying language model. Variations in the underlying language model or changes to the tool's architecture might introduce discrepancies in future evaluations.

\subsection{Discussion}\label{sec:discussion_interface1}
Most experiments were executed completely autonomously, requiring no additional user interaction beyond the initial setup. Researchers only needed to upload their code, provide the execution commands, and download the final reproducibility package. This process eliminates the need for manual environment configuration or the specification of PLs and their versions.

In a few cases, limited user interaction was necessary—typically only to specify a missing dependency. While these interactions were not overly arduous, they highlight an opportunity for further automation. Future improvements could enable the tool to diagnose and resolve such dependency issues automatically without user intervention.

Experiment E7 contained an extensive number of files (approximately 700, including 50 executables), which exceeded the token limit for processing within our current implementation. Our approach extracts the first 50 lines from each file for dependency inference, but in large-scale experiments, this exceeded GPT’s processing constraints. For projects of this scale, the accumulated information exceeds the processing capacity of GPT. Refining our extraction strategy to reduce the volume of data per file may enhance performance on larger projects.

When comparing \tool to other reproducibility tools, our approach demonstrates comparable robustness and adaptability. For example, prior evaluations indicate that tools such as PTU and ReproZip have achieved success rates of up to 94\% (reproducing 17 out of 18 experiments), with Sciunit closely following at 89\% (16 experiments). \toolbackend, developed in previous work~\cite{costa2025backend}, also achieved a 94\% success rate. In contrast, \tool reproduced 15 experiments, corresponding to an 83\% success rate. Overall, these results are encouraging, particularly given that most experiments were executed with little or no interaction.

In the following section, we present a user study comparing \tool with a professional reproducibility platform, \co.

\section{Comparative Study}\label{sec:controlled}
In our previous study, we evaluated \tool’s technical capacity to handle a variety of experiments. In contrast, the present study focuses on assessing its usability and user experience with real researchers. To achieve this, we designed and conducted an empirical study following the methodological recommendations of Wohlin et al.~\cite{Wohlin2012}.

For transparency and reproducibility, all materials related to this study are publicly available. This includes the questionnaires, detailed logs of interactions with 21 participants, and the complete source code of \tool~\cite{costa2025Sciconv}.

\subsection{Study Design}
We are proposing a tool that stands out when compared to the existing state-of-the-art solutions. To assess its effectiveness, it is crucial to compare it against tools already established in the field. Moreover, it is important to evaluate how users perceive its usability in real-world scenarios, as well as the effort required to use it in comparison to current alternatives. To address these objectives, we designed a study that compares our tool with one of the leading platforms on the market, \co.

\subsubsection{Hypotheses}

Informally, we can state our two hypotheses as follows:

\begin{enumerate}
    \item Researchers perceive \tool as having better usability than \co.
    \item Researchers perceive \tool as requiring less effort.
\end{enumerate}

Formally, two hypotheses are being tested: $H_u$ regarding the usability; $H_e$ regarding the effort:

\begin{enumerate}
    \item \textit{Null hypothesis, $H_{u_0}$}: The usability of \tool is equal to the one of \co. $H_{u_0}$: $\mu_d = 0$, where $\mu_d$ is the expected mean of the difference of the usability.
    
    \textit{Alternative hypothesis}: $H_{u_1}$: $\mu_d \neq 0$, i.e., the usability of \tool is different of using \co.

    \item \textit{Null hypothesis, $H_{e_0}$}: The effort of using \tool
    is the same of using \co. $H_{e_0}$: $\mu_d = 0$, where $\mu_d$ is the expected mean of the difference of the effort.
    
    \textit{Alternative hypothesis}: $H_{u_1}$: $\mu_d \neq 0$, i.e., the effort of using \tool different than of using \co.
\end{enumerate}

\subsubsection{Variables}

The independent variables are: for $H_u$ the \textit{SUS score}
and for $H_e$ the \textit{NASA TLX score}.

\subsubsection{Participants and Objects}
We posted information about our study on social media, and we sent messages to people we knew in other research fields so they could share the information about our study. We have also contacted MSc and PhD students in our own institution. We do not offer any compensation except the fact that participants get a reproducibility package for their own experiments.

We used multiple experiments during the study. We used one experiment to provide a tutorial about the two tools involved, another experiment when researchers could not provide their own, and some experiments brought by the researchers themselves to the study.

\subsubsection{Design} 
To evaluate the effectiveness of \tool, we conducted a comparative study against \co, a leading reproducibility platform renowned for its robust performance and user-friendly interface. We employed a within-subjects experimental design, with the single independent factor being the reproducibility tool used and two treatments~\cite{Wohlin2012}: \tool and \co. 

Each participant reproduced the same scientific computational experiment using both tools. This approach minimized the need for participants to familiarize themselves with different experiments, ensuring that both treatments were assessed under similar conditions. To control for potential learning effects, the order of tool exposure was counterbalanced—half of the participants began with \tool, while the other half started with \co.

Additionally, participants were given the option to leverage advanced AI platforms (e.g., ChatGPT, Gemini, etc.) when interacting with \co. This integration allowed us to enhance our evaluation by incorporating state-of-the-art AI support, providing deeper insights into the usability and performance differences between the two solutions.


\subsubsection{Instrumentation}\label{sec:instrumentation}
The study began with each participant reviewing and signing a consent form, followed by completing a pre-study questionnaire\footnote{Available at \url{https://forms.gle/iXY77Jg4Ek4Cagom8}.} to provide demographic information and details about their experience with reproducibility, such as previous attempts and familiarity with relevant tools.

Participants were then assigned to one of two study paths. In the first path, they completed a tutorial on how to use \tool (see \cref{sec:usingSciconv}) with a predefined experiment provided by us. After the tutorial, they attempted to reproduce an experiment independently. Whenever possible, participants were encouraged to bring their own research experiments, using them in both tools to assess their effectiveness in a familiar context. For those without an experiment, we supplied a predefined one to ensure consistency. Once the reproduction process was completed using \tool, participants filled out the SUS questionnaire (SUS questionnaire for \tool\footnote{Available at \url{https://forms.gle/gYgsY6MgJbSi2C3y8}.} and SUS questionnaire for \co\footnote{Available at \url{https://forms.gle/kKA9KiXJzE3FuDfP8}.}) and the NASA TLX questionnaire, both of which are also included in the study’s research artifact~\cite{costa2025Sciconv}. Next, they followed a tutorial on \co (see \cref{sec:usingCodeocean}) and reproduced the same experiment. Afterward, they again completed the SUS and NASA TLX questionnaires, this time evaluating \co.

The second path followed the same structure but in reverse order. Participants initially completed a tutorial on \co, followed by the experiment reproduction and questionnaire assessments. They then switched to \tool, repeating the same steps to ensure a balanced comparison.

At the end of the study, all participants completed a post-study questionnaire\footnote{Available at \url{https://forms.gle/AT5Uf85gosTNZ4Lt6}.} where they provided additional feedback on \tool, including its strengths, weaknesses, and suggestions for improvement.

In summary, a participant would follow the next steps:

\begin{enumerate}
    \item Fill in a consent form
    \item Fill in the pre-questionnaire 
    \item Follow a tutorial on how to use \tool
    \item Reproduce an experiment on \tool
    \item Fill in the SUS questionnaire
    \item Fill in the NASA TLX questionnaire
    \item Follow a tutorial on how to use \co
    \item Reproduce an experiment on \co
    \item Fill in the SUS questionnaire
    \item Fill in the NASA TLX questionnaire
    \item Fill in the post-questionnaire
\end{enumerate}

Half of the participants followed exactly this order, while the other half followed the order (1) (2) (7) (8) (9) (10) (3) (4) (5) (6) (11).


\subsubsection{Data Collection}
We collected data in different forms (SUS questionnaire, the pre-questionnaire, and the post-questionnaire), which we used online Google Forms. For collecting the NASA TLX, we used a web page\footnote{We used the web page provided by Keith Vertanen available at \url{https://www.keithv.com/software/nasatlx/nasatlx.html}.} that mimics the original scale of a possible answer, as well as a similar way to choose between the importance of the different dimensions.

\subsubsection{Analysis Procedure and Evaluation of Vality}

The results of the SUS and the NASA TLX score will be compared pair-wise. Thus, depending on the data distribution, we may use the paired t-test, Wilcoxon, or sign test \cite{Wohlin2012}.

To ensure the validity of the data, one of the authors supported each participant during the entire study. During the study, we were available to clarify any questions and help when participants were stuck. Nevertheless, when answering the questionnaires, participants were left alone so they could provide honest answers.

\subsection{Execution}

To facilitate accessibility, we deployed \tool on a web server, allowing participants to access it from their own computers\footnote{\tool available at \url{http://sciconv.inesctec.pt/}}. 
Since \co is also an online tool, participants could complete the study using any device with a web browser.

To ensure a focused and individualized experience, we conducted the study through one-on-one sessions. Each participant interacted with the tools independently while we observed and provided clarification when needed. To maximize convenience, sessions were scheduled flexibly, allowing participants to join at their preferred time via Zoom.

\subsection{Results}
We now provide the detailed results of the execution of our study.

\subsubsection{Descriptive Statistics}

\paragraph{Subjects}



We were able to execute our study with 21 participants, 12 PhD students, five researchers, three of whom were in industry, and one postdoctoral researcher. 
We had 12 male participants and nine female participants. Almost all participants reported having more than two years of research experience. Specifically, six participants indicated having more than five years but less than ten years of experience, while two participants reported more than ten years of research experience. Notably, only three participants had less than two years of experience. The participants' research areas were diverse, although the majority were related to computing fields such as Software engineering, Machine Learning and Computer Vision, Natural Language Processing/Generation, AI, Data Mining, Information Retrieval, Human–computer interaction (HCI), and Bioinformatics. 
Additionally, there were participants specializing in Biotechnology, Environmental science, Electronics Engineering, and Economics. One participant focused on Construction Digitalization, representing the civil construction area. While almost half (47.6\%) of the participants (10) work in Portugal, others were distributed across several countries, including the USA (3 participants, 14.2\%), France and Germany (2 participants, 9.5\%), and China, Italy, Spain, and India, each represented by one participant (4.8\% each).

\begin{table}[!ht]
    \centering
        \caption{Demographic information of the participants}\rowcolors{2}{white}{gray!25}
    \begin{tabular}
    {>{\raggedright\arraybackslash}m{0.7cm}
>{\centering}m{1.5cm}
>{\centering}m{0.6cm}
>{\centering}m{1.6cm}
>{\centering\arraybackslash}m{2cm}} 
\rowcolor{gray!50}
    \toprule
       \textbf{Id} &  \textbf{Occupation} & \textbf{Gender} & \textbf{Years in research} & \textbf{Research area} \\
         \midrule
P1 & PhD student & Female & $>$ 5 \& $<$ 10  & Software engineering \\
P2 & Researcher & Female & $<$ 2  & Construction digitalization \\
P3 & Industry practitioner & Female & $>$ 2 \& $<$ 5  & Software engineering \\
P4 & PhD student & Female & $>$ 2 \& $<$ 5  & Machine learning and computer vision \\
P5 & Researcher & Male & $<$ 2  &  Information modelling \\
P6 & PhD student & Male & $>$ 2 \& $<$ 5  & Natural language processing/generation \\
P7 & PhD student & Female & $>$ 2 \& $<$ 5  & AI \\
P8 & Industry practitioner & Female & $>$ 10 \& $<$ 20  & Data mining \\
P9 & Industry practitioner & Male & $>$ 10 \& $<$ 20  & Developer \\
P10 & PhD student & Male & $>$ 5 \& $<$ 10  & Information retrieval \\
P11 & PhD student &Male  & $>$ 2 \& $<$ 5  & HCI  \\
P12 & Researcher & Male&  $>$ 5 \& $<$ 10 & Bioinformatics\\
P13 & PhD student & Male &  $>$ 5 \& $<$ 10 & Software engineering \\
P14 & PhD student &Male & $>$ 2 \& $<$ 5  & Economics \\
P15 &PhD student  &Male & $<$ 2   & Biotechnology \\
P16 & PhD student & Male& $>$ 2 \& $<$ 5  & Electronics engineering \\
P17 &Researcher  &Male  & $>$ 2 \& $<$ 5 & Software engineering  \\
P18 &Researcher  &Female &  $>$ 5 \& $<$ 10 & Environmental science \\
P19 &Postdoctoral  & Female &  $>$ 2 \& $<$ 5 & Biotechnology \\
P20 &PhD student & Female & $>$ 5 \& $<$ 10  & Environmental science  \\
P21 & PhD student & Male & $>$ 2 \& $<$ 5  & Software engineering  \\
\bottomrule
    \end{tabular}
    \label{tab:my_label}
\end{table}

Most participants (14 of 21) know the importance of reproducibility in research, though their experience with reproducing experiments varies. One has never tried reproducing software/code, six have attempted fewer than 10 times, and 14 have made over 10 attempts.
However, the familiarity with reproducibility platforms is limited, with 
11 participants were unaware of any platforms. Nine participants have used tools like Binder, ReproZip, Docker, and RenkuLab, but usage varies. Additionally, one participant stated that they used Docker or Anaconda to reproduce their experiments. Reasons for not using these platforms include lack of knowledge, difficulty, or time constraints.

Participants in the study highlighted several challenges related to reproducibility. A common issue reported by several participants was the unavailability of the necessary code (15 participants) or datasets (16 participants), which made it difficult to replicate experiments. 

Additionally, ten participants mentioned software and dependency installation difficulties, as setting up the correct environment, including libraries, proved challenging. Many (16 participants) also faced problems due to missing execution information, where essential instructions or details for running experiments were not provided. While insufficient documentation was mentioned less frequently, one participant still found it a significant barrier.


\rowcolors{2}{white}{gray!25}
\begin{table*}[!ht]
    \centering
    \caption{Participants' experiences with software reproducibility, including their self-rated knowledge (1-5 scale), number of reproducibility attempts, perceived difficulty (1-5 scale), encountered challenges, platforms used, and reasons for not utilizing specific tools}\rowcolors{2}{white}{gray!25}

    \begin{tabular}{llp{1cm}lp{4.6cm}p{2.3cm}p{1cm}p{4cm}}
 \toprule
\rowcolor{gray!50}
\textbf{Id} & \textbf{Know.} & \textbf{\# repro.} & \textbf{Diff.} & \textbf{Challenges} & \textbf{Platforms used} & \textbf{\# used tools} & \textbf{Why not?} \\ 
\midrule

P1 & 3 & 2 & 3 & Code, Data, Missing information & Binder  & 4 &  \\
P2 & 3 & 0 & 2 & Data, Installation, Dependencies & ReproZip  & 0 & I do not have the knowledge to use them \\
P3 & 5 & 10 & 4 & Code, Data, Dependencies, Missing information & None & 5 &  \\
P4 & 5 & 10 & 4 & Data, Installation, Dependencies, Missing information & None & 0 & I do not know them \\
P5 & 1 & 1 & 3 & Installation & None & 0 & I do not know them \\
P6 & 4 & 10 & 3 & Installation, Dependencies, Missing information, Documentation & Docker, Anaconda & 50 &  \\
P7 & 5 & 20 & 5 & Installation, Dependencies, Missing information & None & 0 & I do not know them \\
P8 & 4 & 20 & 4 & Code, Data, Installation, Dependencies, Missing information & Binder  & 5 &  \\
P9 & 5 & 100 & 4 & Code, Data, Installation, Missing information & None & 0 & I do not know them; I did not use them because it would take too long; I did not use them because it was too difficult \\
P10 & 5 & 30 & 3 & Data, Dependencies, Missing information & Binder  & 2 & I tend to pick codes from paper repositories that have a Docker approach for `installation'\\
P11 &4  & 20  & 4 & Code, Data, Missing information & ReproZip  & 5 &  \\
P12 &  5& 50 &5  & Code, Data, Dependencies, Missing information &  Binder, ReproZip, Whole Tale & 10  &  \\
P13 & 5 & 1  &4  & Code, Data, Dependencies, Missing information &  I do not know any & 0 & I do not know them  \\
P14 & 3 & 5 & 3 & Code, Data & I do not know any  & 0 & I do not know them; I have not considered using them  \\
P15 & 3 & 5 & 4 &Code, Data  &  I do not know any & 0  & I do not know them; I did not use them because it was too difficult \\
P16 & 3 & 11 & 4 &Code, Data, Missing information  & Binder, ReproZip  & 2 &  \\
P17 & 4 & 20 & 4 & Code, Data, Dependencies, Missing information &  Binder, Whole Tale & 10 &  \\
P18 & 4 & 5 & 4 & Code, Data, Missing information & I do not know any  & 0 & I did not use them because it was too difficult \\
P19 & 4 & 10 & 4 & Code, Data, Missing information & ReproZip  & 1 &  \\
P20 & 3 & 15 & 3 & Code &  I do not know any & 5 & I do not know them \\
P21 & 5 & 10 & 3 & Code, Dependencies, Missing information &  I do not know any & 0 & I do not know them 
    \label{tab:partici_exp}
\end{tabular}
\end{table*}
\rowcolors{5}{}{}  

\cref{tab:user-experiments} provides an overview of the type of experiment each participant used during the study.

\begin{table}[!ht]
    \centering
    \caption{Experiments used by the participants}\rowcolors{2}{white}{gray!25}

    \begin{tabular}    {>{\raggedright\arraybackslash}m{1cm}
>{\raggedright\arraybackslash}m{7cm}} \rowcolor{gray!50}
\toprule
\textbf{Id} & \textbf{Experiment used} \\
\midrule
P1 & Provided by us \\
P2 & Provided by us \\
P3 & Provided by us \\
P4 & Own experiment using Bash \\
P5 & Own experiment using Python \\
P6 & Own experiment using Python \\
P7 & Own experiment using Bash \\
P8 & Provided by us \\
P9 &  Own experiment using C++ \\
P10 & Provided by us \\
P11 & Provided by us\\
P12 & Provided by us\\
P13 & Own experiment using Python\\
P14 & Own experiment using Bash\\
P15 & Provided by us\\
P16 & Own experiment using Python \\
P17 & Provided by us\\
P18 & Provided by us\\
P19 & Provided by us\\
P20 & Provided by us\\
P21 & Own experiment using Python \\
\bottomrule
    \end{tabular}
    \label{tab:user-experiments}
\end{table}

\paragraph{System Usability Scale}

The possible values range between 0 and 100. The values in
{\color{blue} blue} are associated with \tool, while the {\color{red} red} ones with \co. 

\begin{figure}[!ht]
    \centering
    \includegraphics[width=1\linewidth]{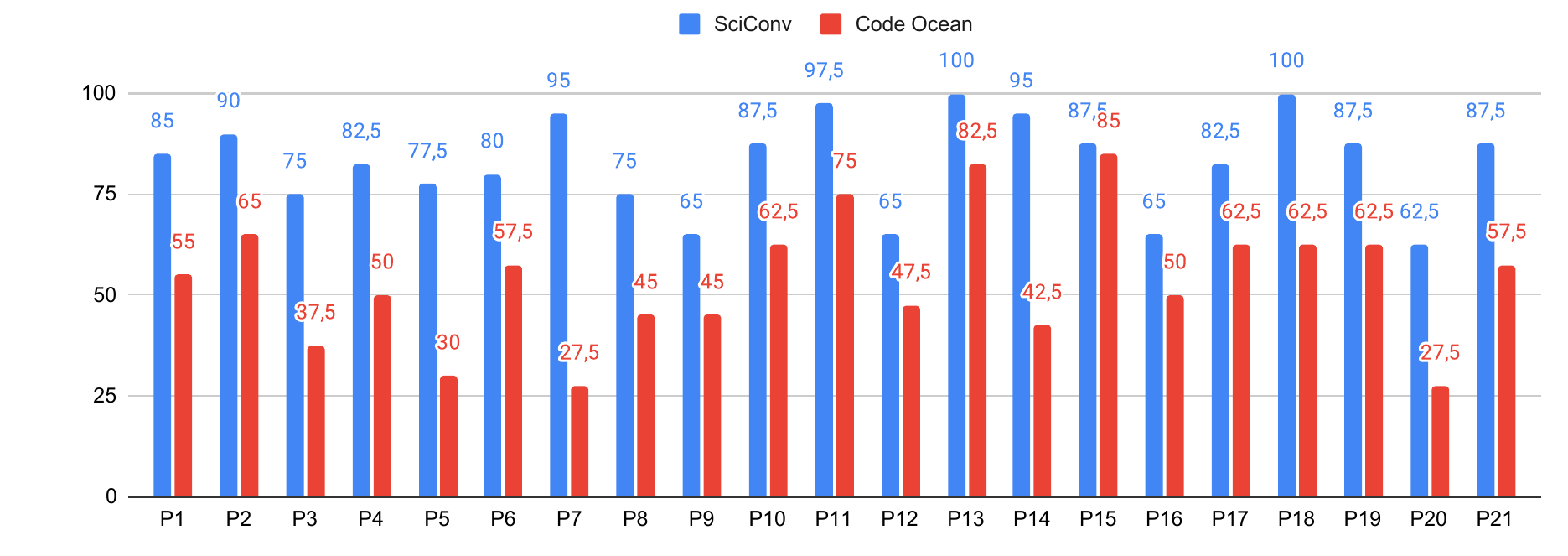}
    \caption{SUS score for each participant and each tool}
    \label{fig:sus}
\end{figure}

\cref{fig:sus} illustrates the SUS score for each participant across both tools, with possible values ranging from 0 to 100. Scores associated with \tool are represented in {\color{blue} blue}, while those corresponding to \co are shown in {\color{red} red}.

\paragraph{NASA TLX}

\begin{figure}[!ht]
    \centering
    \includegraphics[width=\linewidth]{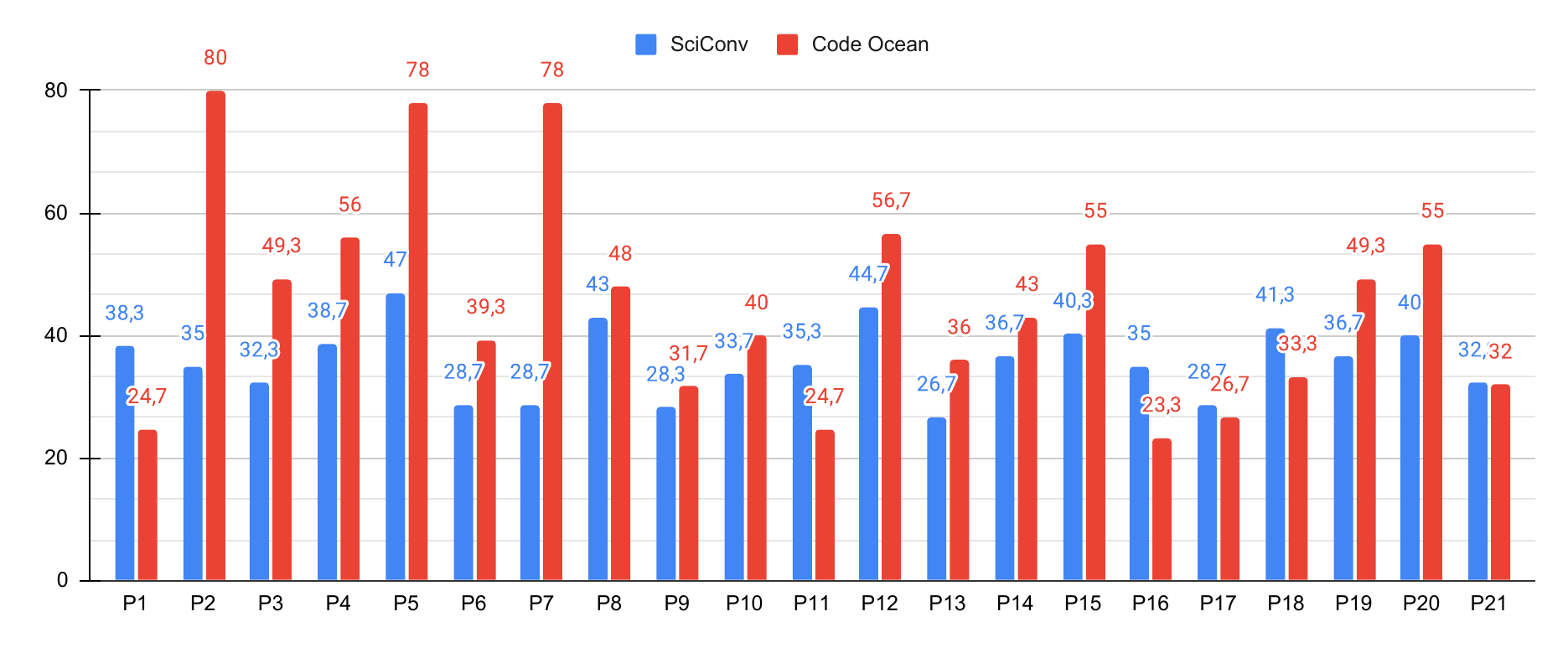}
    \caption{NASA TLX score for each participant and each tool}
    \label{fig:tlx}
\end{figure}

Similarly, \cref{fig:tlx} presents the weighted NASA TLX scores for each participant using both tools, ranging from 0 to 100. As in the previous figure, scores for \tool are depicted in {\color{blue} blue}, while those for \co appear in {\color{red} red}.

\subsubsection{Hypotheses Testing}

We used a Python script and the library \texttt{scipy.stats}\footnote{Available at \url{https://docs.scipy.org/doc/scipy/reference/stats.html}.} to calculate the statistical tests.

\paragraph{Comparing the SUS Scores}
The data did not follow a normal distribution, necessitating the use of a non-parametric test. The p-value computed by the Wilcoxon test \cite{Wohlin2012} is 0.000059, which is less than 0.05, indicating we can reject the null hypothesis $H_{u_0}$ for usability. There is indeed a statistically significant difference between the SUS score for the usage of \tool and \co. 


\paragraph{Comparing the NASA TLX Scores}

The scores for NASA TLX also do not follow a normal distribution. 
Thus, we have also used the Wilcoxon test.
In this case, the p-value is 0.014, which is also less than 0.05. We can thus reject the null hypothesis $H_{e_0}$ for the effort.
The results show that there is a statistically significant difference between the score of the NASA TLX for using \tool and \co.


\subsection{Analysis}

As shown in \cref{fig:sus}, the results show a consistently high score for SUS, never below 62.5 and reaching 100 in two cases, averaging 82.98. These are quite positive numbers, especially compared to \co, which peaked at 85 and was as low as 27.5. The average is 53.81, a little more than half of the average achieved by \tool. The significance of the statistical difference is, in this case, quite evident.

Regarding the workload computed by NASA TLX, we can see similar results in \cref{fig:tlx}. For \tool, the lowest values were 26.7 for one case, 28.3 for one case, and 28.7 for two others, never being higher than 45 in one case; the average was 35.96. Compared to the values achieved by \co, there are some cases with significant differences and some with very small ones. In one case, the minimum was 23.3, even lower than with our tool, but the peak was 80 for one case and 78 for two others, averaging 46.4. There is, however, a statistically significance difference between the two distributions, obviously in favor of \tool.

This is a controlled study, and thus, it cannot be generalized for all researchers.
Nevertheless, the results are quite consistent and promising. Indeed,
participants felt the usability of the conversational interface was greater than that of traditional web-based graphical one. The fact that \tool is able to infer some information required to execute the experiments automatically aids in lowering the workload.

\subsection{Answering RQ: Can a conversational interface improve the usability and accessibility of computational reproducibility tools for researchers across different fields?}\label{sec:RQ1.5-interface}


To address RQ, we designed and developed \tool, a conversational, text-based platform that assists researchers from diverse research disciplines in ensuring computational reproducibility. Our goal was to create a user-friendly software solution that lowers technical barriers, allowing researchers without a computer science background to package and execute computational experiments reliably.

\tool achieves this by leveraging LLM to facilitate natural language interactions, enabling researchers to define computational environments and dependencies with minimal manual input. Unlike traditional reproducibility tools, which often require intricate scripting and deep knowledge of dependency management, \tool automatically infers execution requirements and assists users in resolving issues through an interactive chat-based interface.

To evaluate its effectiveness, we conducted both a qualitative assessment and a comparative user study.
In our qualitative evaluation, \tool successfully reproduced 83\% (15 out of 18) of the experiments from a curated dataset defined in previous work~\cite{Costa2025REP}. The tool automatically inferred execution environments in most cases, requiring minimal interaction. However, limitations were identified, particularly in handling database-dependent experiments and Jupyter Notebook-based workflows.
In the comparative study, \tool was tested against a leading professional reproducibility platform, \co, with 21 researchers from various fields. Results from the SUS and NASA TLX confirmed that \tool significantly outperformed \co in both usability and workload reduction, with statistically significant differences favoring our approach.

\subsection{Threats to Validity}
There are a few threats to the validity of these results, which we divide into four categories as suggested in the literature \cite{Wohlin2012}.

\paragraph{Conclusion Validity} Given the low number of participants, the statistical power is inherently low. Nevertheless, the observed differences are quite evident in almost all cases for SUS and in several cases for NASA TLX. In any case, a bigger number of participants would be preferable.

Issues may occur when collecting data for the SUS and NASA TLX questionnaires. However, the results are mostly consistent among the different researchers, which may indicate no significant problem has occurred.

\paragraph{Internal Validity} We took several measures to ensure the causalities found are not influenced by external factors, except the independent variables.
In particular, since we used a within-subject study, the order by which the participants were exposed to the treatments was alternated, thus minimizing possible learning effects. 

We have also designed the study to have the minimum possible length. Indeed, all participants executed the study in between 45 and 75 minutes.

To collect data, we used mostly Google Forms, which are probably known to many people and thus easy to follow. The web page we used to collect the NASA TLX score was also very easy and intuitive to use.

The experiments used varied among participants. However, the collected results are mostly consistent, which may indicate this factor did not influence the results. However, it is possible that the workload was higher for more complex experiments, thus creating some peaking in the scores. Nevertheless, the experiment used in both tools was the same given the participants, so if such an effect exists, it would affect both tools.

\paragraph{Construct Validity} The tasks we asked participants to do are similar to what they would do for preparing a reproducibility package for their own work. In fact, we asked them to bring their own code whenever possible.
Thus, although we used specific tools to do the task, participants should not have doubts about what they meant. Indeed, several participants showed experience in reproducing experiments (as described in \cref{tab:partici_exp}).

\paragraph{External Validity} The limited number of participants may not represent the broader population of researchers. Most participants are from areas related to computing, spanning various sub-areas such as Software Engineering, Machine Learning, AI, and Natural Language Processing. However, the sample also includes participants from diverse fields like Construction Digitalization, Bioinformatics, Economics, Biotechnology, Electronics Engineering, and Environmental Science, and their results are apparently not different from the others. Notably, participants are distributed across several countries, including Portugal, the USA, France, Germany, Italy, and China, indicating diverse geographical and disciplinary backgrounds. Despite this diversity, they exhibit varying levels of knowledge about reproducibility, ranging from no prior experience to proficient users of reproducibility tools like Binder or more technical solutions like Docker.





\section{Discussion}\label{sec:discussion_interface}
The user study revealed that participants faced significant challenges in selecting the appropriate version of the PL for the experiments they aimed to reproduce. For instance, \co only supports Python versions up to 3.9, which restricts access to the latest language features and libraries. This limitation is not unique to Python, as similar restrictions apply to other PLs, potentially leading to compatibility issues and further complicating the reproducibility process.

Our approach successfully reproduced all the experiments with the correct versions, requiring interaction from only three researchers to resolve an error. However, our approach has limitations that need to be resolved in future iterations. One notable problem is the lack of drag-and-drop functionality for uploading files, which could simplify the user experience, including this feedback: ``The zipped file option has been well received; however, it currently does not allow for the manual addition of files when required.''. Providing a second option for uploading files could increase flexibility and ease of use.

Furthermore, some of the text in our messages requires clarification to enhance user understanding. Feedback from participants indicated that improvements in user interaction are necessary. One participant suggested, ``The interface could be more similar to a chatbot so that we immediately understand that this is like a chatbot.''. Another participant noted, ``Sometimes I do not know that I need to wait for the tool's response, and I think that I need to write something.''.

Participants also recommended implementing color coding for error messages to help users quickly identify when assistance is needed. Additionally, incorporating a loading animation could effectively communicate to users that the chatbot is processing tasks in the background.

These insights highlight the need for ongoing refinement of our platform to improve usability and ensure a smoother interaction process. Addressing these concerns will enhance the overall user experience and contribute to the platform's effectiveness in facilitating reproducible research.

Our approach is a tool for reproducibility providing a UI of interaction to researchers. It offers an intuitive and user-friendly interface that allows researchers to input queries, receive responses, and manage the flow of communication. Thus, principle P8, which we consider a very important one, is now supported by this tool.

\section{Conclusions}\label{sec:conclusions_interface}

This work introduced an innovative conversational tool designed to address the long-standing challenges of computational reproducibility in
research. By leveraging advances in LLM, our approach aims to simplify the reproducibility of experiments, streamline the creation of consistent computational environments, and enhance the overall sharing and verification of scientific results.

Our tool demonstrated significant potential through both qualitative evaluations and comparative user studies. In a qualitative assessment, \tool successfully executed 15 of 18 experiments, achieving an 83\% success rate while highlighting areas for future improvements, such as handling Jupyter Notebooks and database-driven experiments. Comparatively, while other leading tools such as \co, PTU, and ReproZip achieved success rates of up to 94\%, our tool proved competitive, particularly given its conversational and user-friendly interface.

The comparative study further reinforced the value of our tool, showcasing its superior usability and reduced workload compared to \co, a leading professional platform. Statistically significant differences in the SUS and NASA TLX scores underscored its advantages. Participants appreciated the intuitive, chat-based interaction model and the automatic inference capabilities that reduced the need for manual configurations. Feedback from users also highlighted areas for refinement, such as improving feedback messages, introducing drag-and-drop functionality for file uploads, and incorporating visual cues for error handling.

While the current version of \tool demonstrates robust performance, there are clear opportunities for improvement. Expanding support for more complex experiment types, such as those involving user studies experiments or advanced computational setups, will enhance the tool’s applicability across diverse research domains. Furthermore, implementing additional user-requested features and refining interaction dynamics will continue to elevate the user experience.

In summary, our tool represents a significant step forward in addressing the reproducibility crisis in computational science. Combining cutting-edge conversational interfaces with intelligent automation empowers researchers to overcome barriers to reproducibility, ultimately fostering greater trust, transparency, and collaboration within the research community. Moving forward, we aim to enhance \tool’s adaptability to a broader range of experiments, including those with more complex dependencies. Further refinements in interaction design will ensure it remains an accessible and efficient tool for researchers across disciplines.




\balance
\bibliographystyle{IEEEtran}
\bibliography{bibliography}

\appendix

\section{Illustrative Error}\label{app:image}

\begin{figure*}[!htb]
    \centering
    \includegraphics[width=0.95\linewidth]{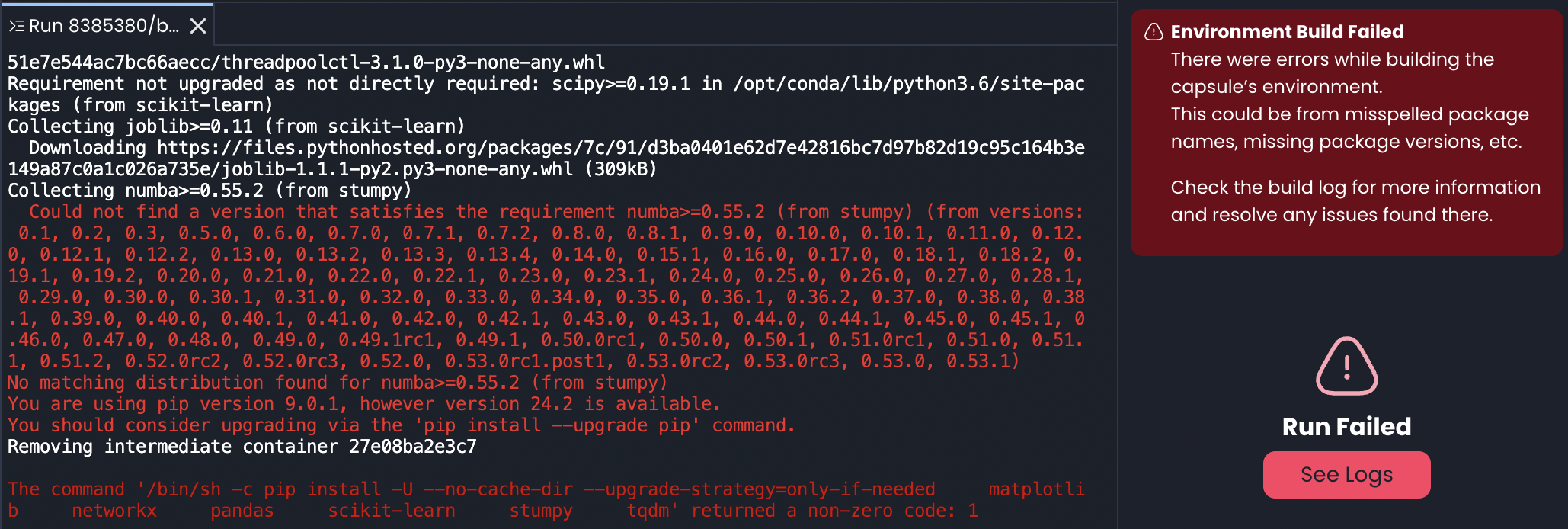}
    \caption{Error shown by Code Ocean when executing an experiment.}
    \label{fig:codeoceanerror}
\end{figure*}

\cref{fig:codeoceanerror} presents the error shown when executing an experiment in Code Ocean. 

\end{document}

%% file: macros.tex
\usepackage{xspace}
\newcommand{\tool}{\textsc{SciConv}\xspace}
\newcommand{\toolbackend}{\textsc{SciRep}\xspace}
\newcommand{\co}{Code Ocean\xspace}
\usepackage{balance}
\usepackage{hyperref}
\usepackage{amsmath}
\usepackage[capitalize,noabbrev]{cleveref}
\usepackage{array}
\usepackage{booktabs}%
\usepackage{graphicx}

\newcommand{\eg}{e.g.\xspace}
\usepackage{colortbl} 
\usepackage{multirow}%
\usepackage{xcolor} 